\documentclass[twocolumn]{aastex631}

\usepackage{amsmath}
\usepackage{verbatim}
\usepackage{multirow}
\usepackage{url}
\usepackage[flushleft]{threeparttable}

\newcommand{\G}{\Gamma}

\newcommand{\sT}{\sigma_{\rm T}}
\newcommand{\p}{^\prime}
\newcommand{\e}{\epsilon}
\newcommand{\g}{\gamma}
\newcommand{\gp}{\gamma^{\prime}}

\newcommand{\ep}{\epsilon^\prime}

\newcommand{\dD}{\delta_{\rm D}}

\newcommand{\psim}{\lower.5ex\hbox{$\; \buildrel \propto \over\sim \;$}}
\newcommand{\lbar}{\lower.0ex\hbox{$\; \buildrel
{\lower0.0ex \hbox{-}} \over\lambda  \;$}}

\newcommand{\cm}{\mathrm{cm}}

\newcommand{\s}{\mathrm{s}}

\definecolor{hookgreen}{rgb}{0.0,0.44,0.0}
\definecolor{periwinkle}{RGB}{104,127,243}
\definecolor{eggplant}{RGB}{83,27,147}
\definecolor{olive}{RGB}{74,80,7}

\shorttitle{TXS 0506+056 in MeV}
\shortauthors{Lewis et al.}

\graphicspath{{./}{Images/}}

\begin{document}

\title{Modeling and Simulations of TXS 0506+056 Neutrino Events in the MeV Band}

\correspondingauthor{Tiffany R Lewis}
\email{tiffanylewisphd@gmail.com}

\author[0000-0002-9854-1432]{Tiffany R. Lewis}
\affiliation{NASA Postdoctoral Program Fellow}
\affiliation{NASA Goddard Space Flight Center\\
Greenbelt, MD 20771, USA}

\author[0000-0002-6774-3111]{Christopher M. Karwin}
\affiliation{Clemson University\\
Clemson, SC 29634, USA}

\author[0000-0002-4188-627X]{Tonia M. Venters}
\affiliation{NASA Goddard Space Flight Center\\
Greenbelt, MD 20771, USA}

\author[0000-0002-0794-8780]{Henrike Fleischhack}
\affiliation{Catholic University of America\\ Washington, DC 20064, USA}
\affiliation{NASA Goddard Space Flight Center\\ Greenbelt, MD 20771, USA}
\affiliation{Center for Research and Exploration in Space Science and Technology, NASA/GSFC\\ Greenbelt, MD 20771, USA}

\author{Yong Sheng}
\affiliation{Clemson University\\
Clemson, SC 29634, USA}

\author[0000-0001-6677-914X]{Carolyn A. Kierans}
\affiliation{NASA Goddard Space Flight Center\\
Greenbelt, MD 20771, USA}

\author[0000-0002-9280-836X]{Regina Caputo}
\affiliation{NASA Goddard Space Flight Center\\
Greenbelt, MD 20771, USA}

\author{Julie McEnery}
\affiliation{NASA Goddard Space Flight Center\\
Greenbelt, MD 20771, USA}

\begin{abstract}


Neutrino detections identified with multiwavelength blazar spectra represent the first $\g$-ray-neutrino multimessenger signals. The blazar, TXS 0506+056 is also unusual in its spectral expressions over time and there are a number of models in the literature which have been proposed. In this work, we model the TXS 0506+056 data during two epochs of neutrino co-observation using a range of Fokker-Planck derived solutions, and simulate the expected response of the proposed AMEGO-X mission for a set of physically plausible scenarios. 

\end{abstract}

\section{Introduction}
\label{Intro}

Blazars are jetted active galaxies with the jet aligned along our line of sight. Due to the beamed relativistic motion of material in the jet, this component dominates the observable spectrum from radio to $\g$-rays. The majority of the observed emission is thought to originate from a specific volume near the base of the jet. The jet environment is generally opaque to radio emission, with specific wavelengths released depending on the distance from the BH. Thus, radio emission is not usually included in blazar models. 

Multiwavelength blazar spectra are usually composed of two broad bumps with a dip in-between. The low-frequency bump is generally recognized as electron synchrotron emission, as confirmed by optical polarization measurements. Various emission mechanisms can contribute to the high-frequency bump, depending on the types of particles being modeled and the target photon fields with which they interact.


BL Lacertae (BLL) type blazars are noted for their lack of optical/UV lines, which is often thought to indicate that the broad emission line region is absent or sufficiently dim that it can be neglected. BLL electron synchrotron emission spectra tend to peak at higher energies (UV to X-rays) and are associated with less powerful jets. In the leptonic picture, the high-frequency bump of a BLL is caused by synchrotron self-Compton (SSC). In the hadronic picture, the high-frequency bump can be caused by proton synchrotron. 

Flat-spectrum radio quasars (FSRQs) have broad emission lines in their optical/UV spectrum, and the electron synchrotron emission tends to peak near the optical band, on average at lower frequencies than BLLs. FSRQs are identified with the higher-powered radio jets of Fanaroff-Riley Type II jets at other orientations. In the leptonic picture, the high-frequency bump of FSRQs is usually attributed to inverse-Compton reprocessing of external photon fields that impinge on the emitting region of the jet, where they interact with the electron(positron) population(s), or external Compton (EC). External photon fields that might impinge on the jet can come from the accretion disk (EC/disk), dust torus (EC/dust), or broad line region (EC/BLR), although in practice, EC/disk is not usually significant. The cosmic microwave background is not expected to contribute significantly to EC components from a blazar primary emitting region that is near the base of the jet. In the hadronic picture, the high-frequency bump of FSRQs may be explained by proton synchrotron, either independently or in combination with synchrotron emission from p$\g$ cascade particle populations.

Neutrinos are the product of hadronic processes. For example, a p$\g$ cascade will produce a neutrino spectrum, and this is one of the more likely avenues to produce observable neutrinos in the blazar environment. A purely leptonic jet cannot produce neutrinos, and a lepto-hadronic jet in which the protons are not sufficiently accelerated to produce observable radiation would be very unlikely to produce an observable neutrino spectrum. As such, definitively identifying neutrinos with a blazar source is the smoking gun for the presence of significant proton acceleration in blazar jets. This is an important observational marker because models that track disk-jet energetics tend to disfavor hadronic and leptohadronic models because they require more energy from the accretion disk than leptonic models, in some cases to the point of seeming unphysical given current accretion models (which is notably an independent area of active study). However, if some blazars definitely produce neutrinos, then we may have to re-examine our understanding of the disk and disk-jet energetics. 

When modeling the emission region in blazar jets, it is important to consider all of the physical processes self-consistently. Particles that are accelerated in one location will tend to emit in the same location, thus we consider the acceleration and cooling regions of the blazar jet to be co-spatial. The minimum variability timescale of a blazar is often understood to indicate the light-crossing timescale of the cooling region that produced the emission. Under this interpretation, it is not unreasonable to model a single homogeneous blob of a size defined by the minimum variability timescale of the high-energy photons. The peak of a flare might be interpreted as the point in time wherein the acceleration and cooling processes could be described by a(n unstable) steady-state. Thus, it follows that a homogeneous, one-zone, steady-state model could represent the physics of the particles in the primary emitting region of a blazar jet. 

The 2014-2015 neutrino flare showed no evidence of multiwavelength flaring activity, and the neutrino flux was $\sim$5 times higher than the average $\gamma$-ray flux, likely implying a strong absorption of GeV photons by an intense X-ray radiation field. This poses a significant difficulty for interpreting the neutrino flares in terms of conventional one-zone models \citep[e.g.][and references therein]{2021ApJ...906...51X}. To overcome this, numerous models that go beyond the framework of the conventional one-zone model have been proposed \citep[e.g.][]{2019PhRvD..99f3008L,2019ApJ...886...23X,2020ApJ...889..118Z,2021ApJ...906...51X}. In this work we focus on the recent two-zone model from \citet{2021ApJ...906...51X}, since it predicts a substantial MeV flux that can be tested with future MeV telescopes such as AMEGO-X.

\citet{2021ApJ...906...51X} consider a two-zone radiation model with an inner and outer blob, where the inner blob is close to the SMBH (within the hot corona), and the outer blob is further away. The two blobs are assumed to be spherical plasmoids of different radii, moving with the same bulk Lorentz factor along the jet axis, and they are filled with uniformly entangled magnetic fields ($\mathrm{B_{in} > B_{out}}$), as well as relativistic electrons and protons. The electrons and protons are assumed to be accelerated primarily by Fermi-type acceleration \citep{2007Ap&SS.309..119R} at the base of the jet, and injected into the blobs. In the inner blob, the X-ray corona provides target photons for efficient neutrino production and strong GeV $\gamma$-ray absorption (which is reprocessed down to the MeV band) resulting from $p \gamma$ interactions. The outer blob is assumed to be away from the broad line region, where it is assumed that relativistic electrons radiate mainly through synchrotron radiation and synchrotron self-Compton scattering. These dissipation processes in the outer blob are responsible for the multiwavelength emission.

Although the inner-outer blob model from \citet{2021ApJ...906...51X} can account for the TXS 0506+056 neutrino flares, it was also shown that the probability of forming a blob within the X-ray corona is likely very small. More generally, however, the corona may still be a promising site for neutrino production via other mechanisms \citep[e.g.][]{2019ApJ...880...40I,2020PhRvL.125a1101M}.


\color{black}
\section{Modeling Methodology}
\label{Motive}

Blazar jets form near the supermassive black hole (SMBH) in the core of active galactic nuclei via a process that is likely related to magnetohydrodynamical winds produced by accretion processes. 
Some of the magnetic field lines from the disk pass through the ergosphere of the BH and bend up (and down) to contain the jet(s) above (below) the plane of the accretion disk. Thus,  material originates from the accretion disk and is funneled into the base of the jet with an energy distribution that is characterized by a thermal distribution with tail.
As the material moves away from the BH, it has a characteristic bulk Lorentz factor $\G = (1-\beta^2)^{-1/2}$, which is determined by the bulk relativistic speed $v = \beta c$, where $c$ is the speed of light. Since the material in a blazar jet axis makes a small angle, $\theta$, with the line-of-sight of the observer, the relativistic Doppler beaming factor $\dD = [\G(1-\beta \cos \theta)]^{-1}$ can be approximated as $\dD \sim \G$. The observed variability timescale, $t_{\rm var}$, constrains the size of the comoving blob via a causality argument regarding the light-crossing timescale. Thus, the radius of the blob in the comoving frame is given by $R'_b \lesssim c\dD t_{\rm var}/(1+z)$, for a cosmological redshift $z$.

In the 
 comoving frame, the electron energy distribution, $N_e(\g)$, is the solution to the steady-state Fokker-Planck equation \citep{lewis18, lewis19}
\begin{align}
\frac{\partial N_e}{\partial t} &= 0 = \frac{\partial^2}{\partial \g^2}\left(\frac{1}{2} \frac{d \sigma^2}{d t} N_e \right) - \frac{\partial}{\partial \g} \left(\left< \frac{d\g}{dt} \right> N_e \right) \nonumber\\
& - \frac{N_e}{t_{\rm esc}} + \dot{N}_{e,{\rm inj}} \delta(\g-\g_{\rm inj}) \ ,
\label{eq-elecTransport}
\end{align}
where $\g \equiv E/(m_ec^2)$ is the electron Lorentz factor, describing the particle energy.  The fourth term on the right-hand side describes the rate of particle injection $\dot{N}_{e,{\rm inj}}$ into the blob with an injection Lorentz factor $\g_{\rm inj}$, which we constrain to be near 1. The third term on the right-hand side describes the energy-dependent (Bohm) diffusive particle escape on timescales given by
\begin{equation}
t_{\rm esc}(\g) = \frac{\tau}{D_0 \g} = \frac{R^{\prime 2}_b q B D_0}{m_e c^3} \ ,
\end{equation}
where $\tau$ is a dimensionless escape parameter that depends on the comoving radius of the blob, the fundamental charge, the jet magnetic field, the electron mass, and the stochastic diffusion parameter $D_0 \propto s^{-1}$ \citep{park95}.


Equation (\ref{eq-elecTransport}) includes a broadening coefficient term,
\begin{equation}
\frac{1}{2} \frac{d \sigma^2}{d t} = D_0 \g^2\ ,
\end{equation}
which encodes a formulaic description of second-order Fermi (stochastic) acceleration and includes the first term of the drift coefficient. We assume the hard-sphere approximation for characterizing the stochastic acceleration. The drift coefficient is given by
\begin{equation}
\left< \frac{d\g}{dt} \right> = D_0\left[ 4\g + a\g - b_{\rm syn}\g^2 - \g^2 \sum_{j=1}^J b_C^{(j)} H(\g\epsilon_{\rm ph}^{(j)})\right] \ ,
\end{equation}
and includes terms (from left to right) for first-order (shock) acceleration, adiabatic expansion with parameter $a$, synchrotron cooling with parameter $b_{\rm syn}$, and Compton cooling with parameter $b_{\rm C}^{(j)}H(\g\epsilon_{\rm ph}^{(j)})$. The Compton cooling term includes contributions from all of the relevant photon fields denoted by $(j)$ with Klein-Nishina effects encoded in the $H(\g\epsilon_{\rm ph}^{(j)})$ term determined by the characteristic energy of the photon field $\epsilon_{\rm ph}^{(j)}$ \citep[for additional details see][]{lewis18}.


The evolution of the proton energy distribution is described by a similar Fokker-Planck equation to that for the electrons, except with an additional term,
%
%
\begin{align}
\frac{\partial N_p}{\partial t} &= 0 = \frac{\partial^2}{\partial \g^2}\left(\frac{1}{2} \frac{d \sigma^2}{d t} N_p \right) - \frac{\partial}{\partial \g} \left(\left< \frac{d\g}{dt} \right> N_p \right) \nonumber\\
& - \frac{N_p}{t_{\rm esc}} - D_0 f_{\rm cascade}(\g) N_p + \dot{N}_{p,{\rm inj}} \delta(\g-\g_{\rm inj}) \ .
\label{eq-protTransport}
\end{align}
The additional $f_{\rm cascade}(\g)$ term 
%
represents the protons that leave the system because they become other particles arising from hadronic interactions. The loss of protons due to these interactions is 
%
energy dependent and calculated in PYTHIA 8~\citep{PYTHIA8p2}.\footnote{Additional details to be provided in an upcoming publication in preparation.} 
The escape and broadening coefficient terms in the proton Fokker-Planck equation are the same as those in the electron Fokker-Planck equation, except for mass corrections where applicable. The proton drift coefficient is similar to the electron drift coefficient,
\begin{equation}
\left< \frac{d\g}{dt} \right> = D_0\left[ 4\g + a\g - b_{\rm syn}\g^2 - b_{p\g}(\epsilon^*) \g \right] \ ,
\end{equation}
except that we have excluded the term related to cooling via Compton scattering, which is not expected to be relevant for the proton population, and included a cooling term due to the $p\g$ interactions. 
%
The $p\g$ cooling coefficient is defined in  \citet{boett13} as
\begin{equation}
b_{p\g}(\epsilon^*) \equiv \frac{c \left<\sigma_{p\g} f_{p\g}\right>}{D_0} n_{\rm ph}(\epsilon^*) \epsilon^* \ ,
\end{equation}
where $\epsilon^* \sim 0.95 [{\rm erg}]/(E_p[{\rm erg}])$ is a dimensionless energy that depends on the proton energy $E_p$ in ergs, and $\left<\sigma_{p\g} f_{p\g}\right> \approx 10^{-28}$ cm is the elasticity-weighted interaction cross-section for the $\Delta$ resonance, which is expected to be the most relevant hadronic interaction process for blazars \citep{mucke00}.

The spectral emission components include direct emission from the accretion disk \citep{shakura73}
and dust torus, in addition to emission from the jet due to synchrotron,
SSC, and EC of dust torus and BLR photons in the leptonic case and electron synchrotron, SSC and EC/dust, in addition to proton synchrotron and cascade emission in the leptohadronic case.  The blazar TXS 0506+056 has
redshift $z=0.3365$ giving it a luminosity distance
$d_L=5.5\times10^{27}\ \cm$ in a cosmology where $(h, \Omega_m,
\Omega_\Lambda) = (0.7, 0.3, 0.7)$.

The $\nu F_\nu$ disk flux is approximated as
\begin{flalign}
f^{\rm disk}_{\e_{\rm obs}} = \frac{1.12}{4\pi d_L^2}\ 
\left( \frac{ \e}{ \e_{\rm max}} \right)^{4/3} {\rm e}^{-\e / \e_{\rm max}}
\end{flalign}
\citep{dermer14} where $\e=\e_{\rm obs}(1+z)$ and $m_e c^2 \e_{\rm max} = 10$ eV.
The $\nu F_\nu$ dust torus flux is approximated as a blackbody, 
\begin{flalign}
f^{\rm dust}_{\e_{\rm obs}} = 
\frac{15 L^{\rm dust}}{4\pi^5d_L^2} \frac{(\e/\Theta)^4}{\exp(\e/\Theta) - 1} \ ,
\end{flalign}
where $T_{\rm dust}$ is the dust temperature, $\Theta = k_{\rm B}T_{\rm dust}/(m_ec^2)$, and $k_B$ is the Boltzmann constant.

The primary emission region $\nu F_\nu$ flux is computed using the particle distributions, which are the solutions to
the electron and proton Fokker-Planck equations (Equation [\ref{eq-elecTransport} \& \ref{eq-protTransport}]), 
$N\p_e(\gp)$ \& $N\p_p(\gp)$ respectively.  We now add primes to indicate the distribution is in the
frame co-moving with the blob.

The electron synchrotron flux
\begin{flalign}
f_{\epsilon_{\rm obs}}^{e,{\rm syn}} = 
\frac{\sqrt{3} \e' \delta_{\rm D}^4 q^3 B}{4\pi h d_{\rm L}^2} 
\int^\infty_1 d\gp\ N\p_e(\gp)\ R(x_e)\ ,
\label{eq-fsyn-e}
\end{flalign}
where 
\begin{flalign}
x_e = \frac{4\pi \epsilon' m_e^2 c^3}{3qBh\g^{\prime 2}}\ ,
\label{eq-xe}
\end{flalign}
and $R(x)$ is defined by \citet{crusius86}.  Similarly, the proton synchrotron flux
\begin{flalign}
f_{\epsilon_{\rm obs}}^{p,{\rm syn}} = 
\frac{\sqrt{3} \e' \delta_{\rm D}^4 q^3 B}{4\pi h d_{\rm L}^2} 
\int^\infty_1 d\gp\ N\p_p(\gp)\ R(x_p)\ ,
\label{eq-fsyn-p}
\end{flalign}
where 
\begin{flalign}
x_p = \frac{4\pi \epsilon' m_p^2 c^3}{3qBh\g^{\prime 2}}\ ,
\label{eq-xp}
\end{flalign}
with mass corrections and different particle distributions, but notably no other parameter changes.  Synchrotron self-absorption is also calculated for each case.

The SSC flux
\begin{flalign}
f_{\e_s}^{\rm SSC} & = \frac{9}{16} \frac{(1+z)^2 \sT \e_s^{\prime 2}}
{\pi \dD^2 c^2t_{v}^2 } 
 \int^\infty_0\ d\ep_*\ 
\frac{f_{\e_*}^{\rm syn}}{\e_*^{\prime 3}}\ 
\nonumber \\ & \times
\int^{\infty}_{\gp_{1}}\ d\gp\ 
\frac{N\p_e(\gp)}{\g^{\prime2}} F_{\rm C}\left(4 \g\p \ep_{*}, \frac{\e}{\g\p} \right) \ ,
\label{fSSC}
\end{flalign}
\citep[e.g.,][]{finke08_SSC} where $\ep_s= \e_s(1+z)/\dD$, $\ep_*=\e_*(1+z)/\dD$, and
\begin{flalign}
\gp_{1} = \frac{1}{2}\epsilon\p_s 
\left( 1+ \sqrt{1+ \frac{1}{\epsilon\p \epsilon\p_s}} \right) \ .
\end{flalign}
The function $F_{\rm C}(p,q)$ was originally derived by
\citet{jones68}, but had a mistake that was corrected by
\citet{blumen70}.  The EC flux \citep[e.g.,][]{georgan01,dermer09}
\begin{flalign}
\label{eq-ECflux}
f_{\e_s}^{\rm EC} & = \frac{3}{4} \frac{c\sT \e_s^2}{4\pi d_L^2}\frac{u_*}{\e_*^2} \dD^3 
\nonumber \\ & \times
\int_{\g_{1}}^{\g_{\max}} d\g
\frac{N\p_e(\g/\dD)}{\g^2}F_{\rm C} \left(4 \g \e_*, \frac{\e_s}{\g} \right) \ ,
\end{flalign}
where the lower limit is
\begin{flalign}
\g_{1} = \frac{1}{2}\e_s \left( 1+ \sqrt{1+ \frac{1}{\e \e_s}} \right) \ ,
\end{flalign}
the energy density $u_*$ and dimensionless photon energy $\e_*$ describe the external radiation field.  The energy density of the dust torus photon field, 
\begin{equation}
u_* = u_{\rm dust} = 2.2 \times 10^{-5}  \bigg(\frac{ \xi_{\rm dust}}{0.1} \bigg) 
\bigg( \frac{T_{\rm dust}}{1000\ {\rm K}} \bigg)^{5.2}  ~ {\rm erg ~ cm^{-3}} \ 
\end{equation}
and the corresponding dimensionless energy
\begin{equation}
\e_* = \e_{\rm dust} = 5 \times 10^{-7} 
\bigg( \frac{T_{\rm dust}}{1000\ {\rm K}} \bigg) \ ,
\end{equation}
are consistent with \citet{nenkova08-pt2}.  We note that the dust reprocessing efficiency $\xi_{\rm dust}$ is a free parameter.  The energy density of the BLR photons takes the form, 
\begin{equation}
u_* = u_{\rm line} = \frac{u_{\rm line,0}}{1 + (r_{\rm blob}/r_{\rm line})^\beta} \ ,
\label{eq-uline}
\end{equation}
where  $r_{\rm blob}$ is the
distance of the emitting blob from the black hole (a free parameter) and $\beta\approx 7.7$ \citep{finke16}.
The line radii $r_{\rm line}$ and intrinsic energy densities $u_{\rm line,0}$ for all broad lines used are known ratios based on composite SDSS quasar spectra \citep[e.g.][]{vanden01, finke16}.  The parameters $r_{\rm
  line}$ and $u_{\rm line,0}$ are determined from the disk luminosity
using relations found from reverberation mapping \citep{finke16,lewis18}

\color{black}

\color{black}
\section{Data Analysis}
\label{Data}
\subsection{Historical Spectra}
\label{history}

Since the data available during the periods of interest is limited, especially in particular wavelengths, it is prudent to first apply the model to more complete, although non-simultaneous historical data for the source. This helps to calibrate the model and provide a baseline for the parameter space that might be specific to this source. Notably, the comparison between the model and non-simultaneous data should not be understood as an explanation of the historical activity of the source since the historical data spans many distinct states of the blazar's activity. Each state might be driven by specific physical processes, and the average of all of these observations is not necessarily related to the specifics of any physical state.  On a related note, since the comparison to historical data is only for the purpose of calibration for the free parameters, it is not especially interesting to perform the comparison for all of the model types used to fit particular flares. Therefore, we used only the leptohadronic model where the SSC component is not included in the electron transport equation (the FSRQ interpretation). 

The historical data was acquired from the Space Science Data Center (SSDC) SED Builder tool, which provides broadband spectral and light curve data from a selection of previously published data sets \citep{Myers03, Healey07, Jackson07, Nieppola07, Condon98, Wright94, Planck11, Gregory96, White92, Planck14, Planck15, Wright10, Bianchi11, Evans14, Voges99, Boller16, Abdo10, Nolan12, Acero15, Bartoli13}. In the radio through UV, the data spans roughly 30 years. The radio observations were made by AT, CLASSSCAT, CRATES, GBT, JVASPOL, NIEPPOCAT, NVSS, PMN, and Planck. The IR data was observed by WISE, and the Opt/UV data were observed by GALEX and Swift.  The X-rays were observed by ROSAT and {\it Swift}. The $\g$-ray data from Fermi{\it LAT} is limited to the 3FGL and 2FHL catalogues since the current catalogue versions include all historical measurements of the source (avoiding double counting of older data), and previous versions of those catalogues tend not to include updated diffuse background models and analysis methods (making them less accurate). There is one upper limit from ARGO2LAC in the very high-energy $\g$-rays. 

\begin{figure}[t]
\centering
\includegraphics[width=0.48\textwidth,trim={0 0 0 18},clip]{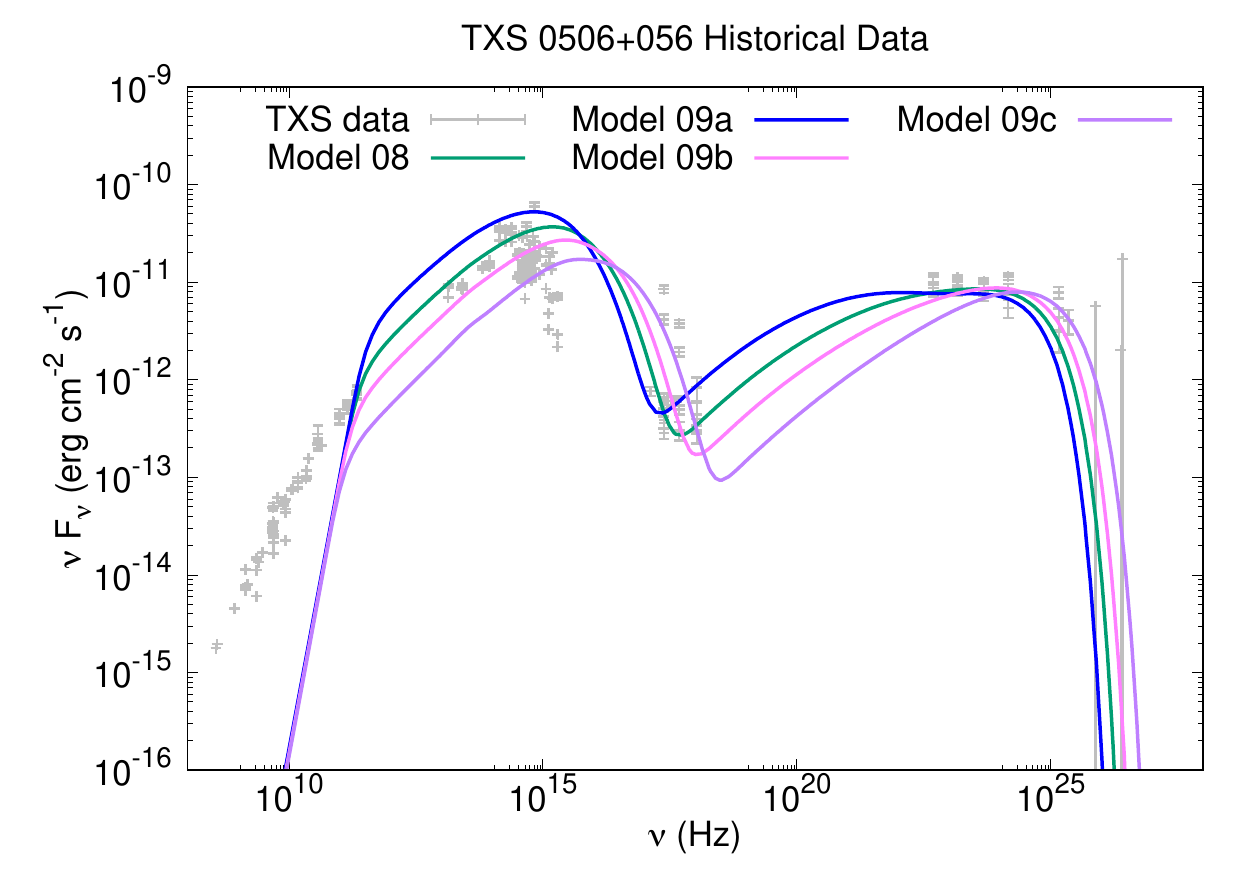}
\caption{A range of leptohadronic model parameterizations are compared to historical data to constrain the baseline parameter space for TXS 0506+056. It is only necessary to change a single parameter to illustrate this range of data.  Model 09a is represented in blue and has a magnetic field of $B=1.2$G. Model 08 (green) has $B=0.9$G.  Model 09b (pink) has $B=0.7$G. Model 09c (purple) has $B=0.5$G.} 
\label{fig-TXShist}
\end{figure}

The range of models in comparison to the historical data is produced by varying the strength of the magnetic field from $0.5$ to $1.2$ G (Table \ref{tbl-freeparams}). The remainder of the free parameters were varied during the analysis, but are held constant through the demonstration of possible parameterizations of the historical data of TXS 0506+056, shown in Figure \ref{fig-TXShist}. Model 09a (blue) has the highest magnetic field value (1.2 G), which produces a synchrotron peak at lower frequencies and with a higher flux, consistent with the highest UV data. The break between the high and low frequency bumps occurs in the softer X-rays, consistent with the lower flux cluster of soft X-rays.  While the combination of SSC and EC components replicates the flat $\g$-ray spectrum and sharp cutoff at higher frequencies, the frequency of the cutoff is not well matched.

Model 09c (purple in Figure \ref{fig-TXShist}) has the smallest value for the magnetic field $B=0.5$G. Here, the electron synchrotron curve is less luminous and peaks at higher frequencies. Since the Compton emission is also at higher frequencies, the convergence of the low and high frequency bumps is at a higher frequency. In this case, the higher X-ray data would be explained as the high-frequency side of the synchrotron curve, rather than any physical component having higher flux as one might expect of a flare. Magnetic fields can suddenly decrease during magnetic reconnection. The peak of the Compton curve in this case explains the highest frequency $\g$-ray data, and turns over to respect the VHE upper limits. 

Most of the historical data can be explained within the range of parameters considered. These models are not unique representations of the possible physics that occurred in this blazar over a 30 year period. However, the parameters provide a baseline for more specific analysis of simultaneous data with this family of transport models.  Even though the models consider the acceleration and emission processes of both electrons and protons (including relevant cascades), the electrons and protons are constrained to be equal in number at the rate of injection. The natural consequence of this is that the proton synchrotron emission component is subdominant, and the leptonic emission processes form the observed spectrum at all wavelengths.

\begin{deluxetable*}{lrrrrr}
\tablecaption{Free Model Parameters \label{tbl-freeparams}}
\tablewidth{\textwidth}
\tablehead{
\colhead{Parameter (Unit)}
& \colhead{Historical}
& \colhead{FSRQ 2014}
& \colhead{FSRQ 2017}
& \colhead{BLL 2014}
& \colhead{BLL 2017}
}

\startdata
$t_{\rm var}$ (s)
&$1.0 \times 10^{4}$
&$8.0 \times 10^{3}$
&$1.0 \times 10^{3}$
&$9.0 \times 10^{2}$
&$5.0 \times 10^{3}$
\\
$B$ (G)
&$0.5, \, 0.7, \, 0.9, \, 1.2$
&$1.1$
&$1.05$
&$0.5$
&$3.0$
\\
$\delta_{\rm D}$ 
&$35$
&$40$
&$72$
&$100$
&$20$
\\
$r_{\rm blob}$ (cm)
&$4.0 \times 10^{17}$
&$4.0 \times 10^{17}$
&$6.2 \times 10^{17}$
&$9.7 \times 10^{17}$
&$1.0 \times 10^{18}$
\\
$\xi_{\rm dust}$ 
&$0.1$
&$0.1$
&$0.1$
&$0.001$
&$0.1$
\\
$T_{\rm dust}$ (K)
&$600$ 
&$600$ 
&$800$ 
&$300$ 
&$800$
\\
$L_{\rm disk}$ (erg s$^{-1}$)
&$1.0 \times 10^{45}$ 
&$1.0 \times 10^{45}$ 
&$4.0 \times 10^{45}$ 
&$1.0 \times 10^{45}$ 
&$1.0 \times 10^{46}$
\\
$D_0$ (s$^{-1}$)
&$4.8\times 10^{-6}$
&$4.8\times 10^{-6}$
&$8.0\times 10^{-6}$
&$7.0 \times 10^{-6}$
&$1.5 \times 10^{-3}$
\\
$a$
&$-3.9$
&$-3.9$
&$-3.9$
&$-3.9$
&$-3.5$
\\
$\gamma_{\rm inj}$
&$1.01$
&$1.01$
&$1.01$
&$1.01$
&$1.01$
\\
$L_{\rm inj}$ (erg s$^{-1}$)
&$6.6 \times 10^{28}$
&$3.3 \times 10^{28}$
&$9.4 \times 10^{28}$
&$7.4 \times 10^{28}$
&$8.3 \times 10^{28}$
\enddata
\end{deluxetable*}

\begin{deluxetable*}{llllllr}
\tablecaption{Calculated Parameters \label{tbl-calculated}}
\tablewidth{0pt}
\tablehead{
\colhead{Parameter (Unit)}
& \colhead{Historical}
& \colhead{FSRQ 2014}
& \colhead{FSRQ 2017}
& \colhead{BLL 2014}
& \colhead{BLL 2017}
}

\startdata 
$R_{Ly\alpha}$ (cm)
&$2.7 \times 10^{16}$ 
&$2.7 \times 10^{16}$  
&$5.7 \times 10^{16}$ 
&$2.7 \times 10^{16}$ 
&$9.4 \times 10^{16}$ 
\\
$R_{H\beta}$ (cm)
&$1.0 \times 10^{17}$ 
&$1.0 \times 10^{17}$ 
&$2.2 \times 10^{17}$ 
&$1.0 \times 10^{17}$ 
&$3.5 \times 10^{17}$ 
\\ 
$R\p_{b}$ (cm)
&$7.9 \times 10^{15}$
&$7.2 \times 10^{15}$
&$1.6 \times 10^{15}$
&$2.0 \times 10^{15}$
&$2.2 \times 10^{15}$
\\
$\phi_{j,{\rm min}}$ ($^{\circ}$)
&$0.6$
&$0.5$
&$0.07$
&$0.01$
&$0.01$
\\
\hline
\\
$P_B$ (erg s$^{-1}$)
&$1.4-8.2 \times 10^{44}$
&$7.5 \times 10^{44}$
&$1.1 \times 10^{44}$
&$7.6 \times 10^{43}$
&$1.4 \times 10^{44}$
\\
$P_{e+p}$ (erg s$^{-1}$)
&$1.3-10 \times 10^{44}$
&$7.3 \times 10^{44}$
&$1.5 \times 10^{45}$
&$1.3 \times 10^{45}$
&$3.9 \times 10^{44}$
\\
$\zeta_{e+p}^{\dagger}$
&$1.6-3.9$
&$1.0$
&$13$
&$17$
&$2.9$
\\
$\mu_{a}^{\ddagger}$
&$0.3-0.9$
&$0.6$
&$0.2$
&$0.6$
&$0.02$
\\
\hline
\\
$u_{\rm ext}$ (erg cm$^{-3}$)
&$2.1\times 10^{-5}$
&$2.1\times 10^{-5}$
&$6.2\times 10^{-5}$
&$5.9\times 10^{-8}$
&$6.4\times 10^{-5}$
\\
$u_{\rm dust}$ (erg cm$^{-3}$)
&$1.5 \times 10^{-6}$ 
&$1.5 \times 10^{-6}$ 
&$6.9 \times 10^{-6}$ 
&$4.2 \times 10^{-10}$ 
&$6.9 \times 10^{-6}$ 
\\
$u_{\rm BLR}$ (erg cm$^{-3}$)
&$2.0 \times 10^{-5}$
&$2.0 \times 10^{-5}$
&$5.5 \times 10^{-5}$
&$5.8 \times 10^{-8}$
&$5.7 \times 10^{-5}$
\\
\hline
\\
$A_C$
&$0.5-2.6$
&$0.7$
&$7.3$
&$0.6$
&$0.07$
\\
$L_{\rm jet}$ (erg s$^{-1}$)
&$2.5-6.3 \times 10^{43}$
&$2.9 \times 10^{43}$
&$2.4 \times 10^{43}$
&$5.9 \times 10^{42}$
&$3.6 \times 10^{44}$
\\
$\sigma_{\rm max}$
&$3.8$
&$3.5$
&$1.3$
&$1.4$
&$340$
\\
\enddata
\tablenotetext{\dagger}{ $\zeta_{e+p} = (u_e+u_{p})/u_B = (P_e + P_{p})/P_B$ is the equipartition parameter, where $\zeta_{e+p} = 1$ indicates equipartition between the particles and field.}
\tablenotetext{\ddagger}{ $\mu_a \equiv (P_B+P_e+P_p)/P_a$ is the ratio of jet to accretion power.  Magnetically arrested accretion explains values of $\mu_a \lesssim$ a few. } 
\end{deluxetable*}

\color{black}
\subsection{2014/15 Epoch}
\label{2014}

\begin{figure}[t]
\centering
\includegraphics[width=0.47\textwidth, trim={4 0 4 18}, clip]{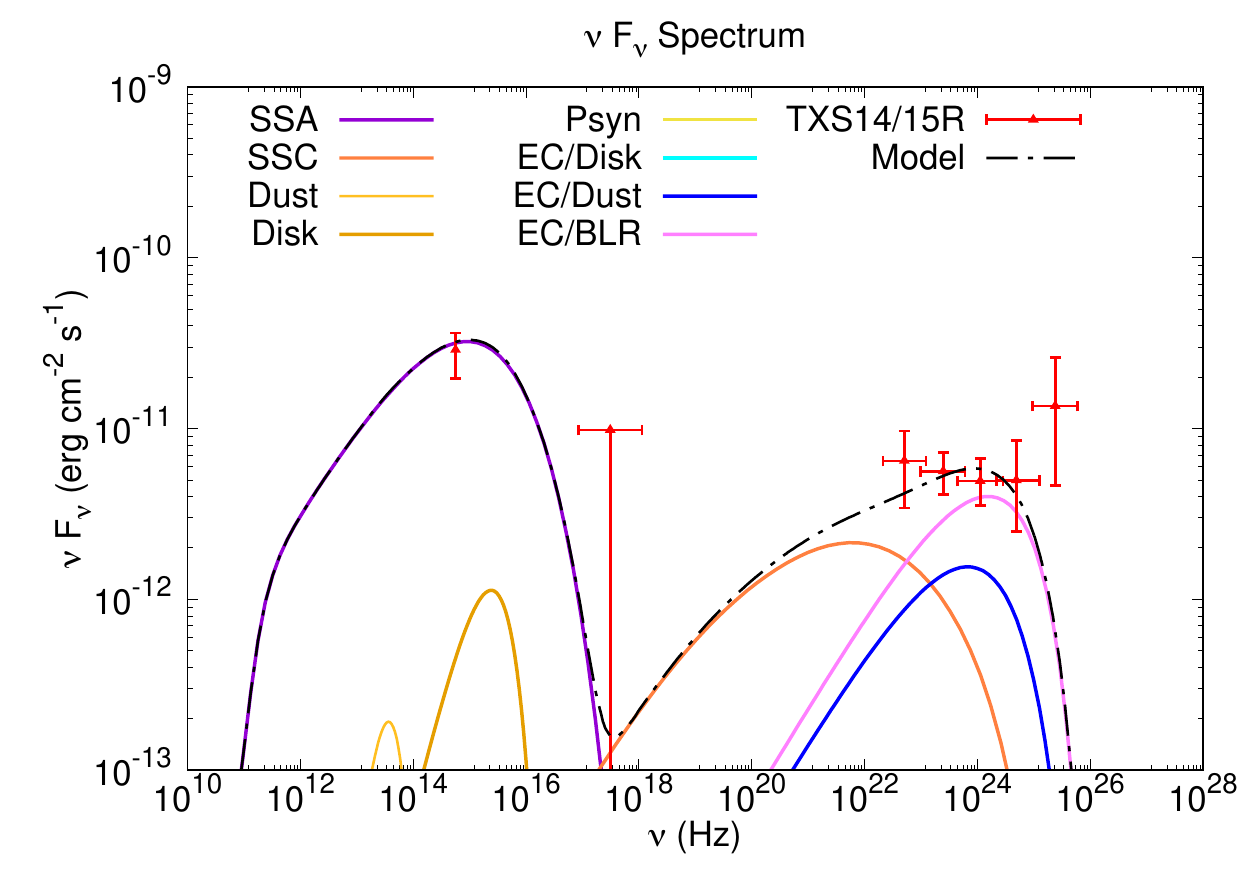}
\caption{Comparison of the FSRQ leptohadronic model to broadband multiwavelngth data from the 2014-2015 flare of TXS 0506+056 \citep[][in red]{Rodrigues19}. The individual components of the emission model are labeled with the EC components dominating in the $\g$-rays. The co-added total emission from the model is given by the dot dashed line.}
\label{fig-flare14no}
\end{figure}

The broadband spectral data for the 2014/2015 flare  \citep{Rodrigues19} is not dissimilar to the historical range for this source. This is reflected in the model comparison with the 2014 flare in Figure~\ref{fig-flare14no}, where we use the leptohadronic model without the SSC cooling term in the electron transport equation. This model is valid where the SSC component is subdominant compared with the EC emission, and can therefore be neglected, which is the case for FSRQ. While TXS 0506+056 has historically been classified as a LSP BL Lac, some debate has arisen in recent years, and we here examine the parameters of a simulation that intentionally emphasizes the EC over the SSC component. Since this is the model we calibrated with the historical data, we should expect that a lot of the parameters will be similar, even though the available simultaneous data in this case is limited. 

In Figure \ref{fig-flare14no}, the optical data is well explained by the synchrotron peak, and the X-ray upper limit is well respected. The concave nature of the $\g$-ray data (which may be an upward statistical fluctuation) is not possible to recreate in a model with only one dominant physical component in that regime. In particular, the highest frequency data point cannot be matched while constraining the model in this way. 
In this analysis, we assume that TXS 0506+056 behaves like an FSRQ during the 2014 flare (which is not necessarily expected due to the low Compton dominance $A_C$ (Table \ref{tbl-calculated}).

\begin{figure}[t]
\centering
\includegraphics[width=0.47\textwidth, trim={4 0 4 0}, clip]{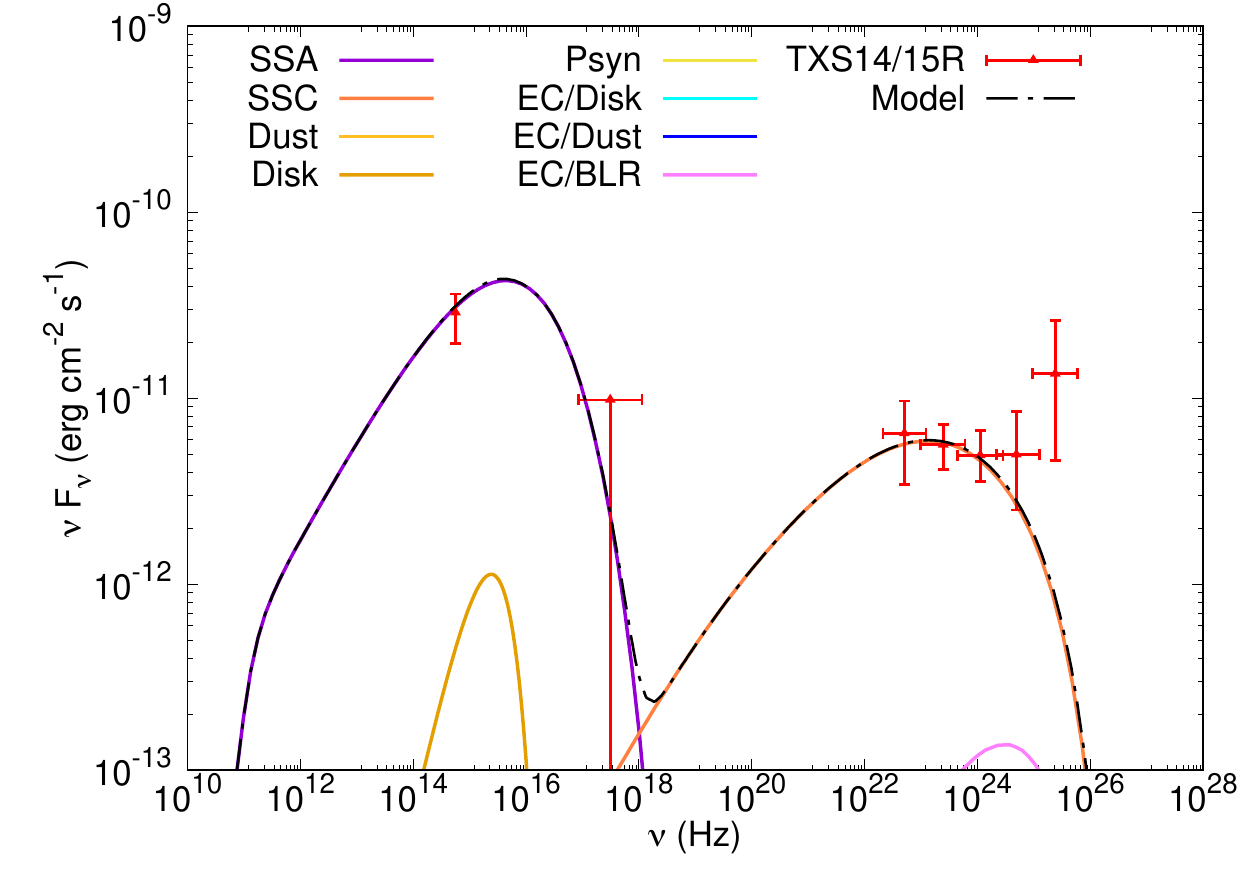}
\caption{Comparison of the BLL leptohadronic model to broadband multiwavelngth data from the 2014-2015 flare of TXS 0506+056 \citep[][in red]{Rodrigues19}. The individual components of the emission model are labeled with the SSC component dominating in the $\g$-rays. The co-added total emission from the model is given by the dot dashed line.}
\label{fig-flare14ssc}
\end{figure}

Figure \ref{fig-flare14ssc} explores the same data set \citep{Rodrigues19}, under the assumption that TXS 0506+056 is a BLL. In this case, the electron transport equation includes SSC as a cooling term, and the parameters are tuned intentionally to emphasize the SSC component rather than EC.  The optical data is well explained with electron synchrotron emission. While the synchrotron peak occurs much closer to the X-ray upper limit, there is no violation of it. The fit to the $\g$-ray data seems more natural in this configuration, with the SSC component following the shape of the four lowest frequency points. However the highest frequency $\g$-ray point remains unexplained under the paradigm of of single dominant component for the high-energy bump. In order to suppress the EC/BLR, the blob is further from the BH to decrease the energy density of the photon fields incident on the jet from the BLR.  In order to suppress the EC/Dust component, it was insufficient to reduce the temperature of the dust torus, and instead we reduce the reprocessing efficiency $\xi_{\rm dust}$ by 2 dex.

In the FSRQ case, the magnetic field $B=1.1$G is on the higher side of the range examined for the historical data, while in the BLL case it is at the lower end $B=0.5$G. The minimum variability timescale, which may be interpreted as the light crossing timescale in the emission region, $t_{\rm var} = 8 \times 10^3$s is slightly smaller than the historical case for the FSRQ and much smaller in the BLL case. From that it also follows that the size of the emitting region in the comoving frame $R'_b$ is slightly smaller or much smaller, respectively. Since the bulk Lorentz factor $\G = \dD$ is higher in the FSRQ case while the acceleration and emission processes occur at the same rate as the historical case, the injected particle luminosity $L_{\rm inj}$ can be lower.  In the BLL case, there is more acceleration generated from stochastic scattering ($D_0$), but also more energy lost to adiabatic expansion, synchrotron radiation, and Compton emission (since the SSC component is broader). All together, this creates the need for a higher injection luminosity in the BLL case, as well as more Doppler boosting into the observer frame to explain the same data. 

The Distance of the emitting region from the BH $r_{\rm blob}$ (Table \ref{tbl-freeparams}) is well under parsec-scale in all cases where the leptonic emission dominates. In the FSRQ case, the emitting region (blob) sits just outside the outer edge of the broad line region, which is roughly signified by the distance of the H$\beta$ line from the BH $R_{H\beta}$ (Table \ref{tbl-calculated}).  In the BLL case, the model indicates that the emitting region is over twice the distance from the BH and well outside the BLR, which is one reason that we might observe less EC emission - the energy density of the external photon fields drop with distance. The jet opening angle $\phi_{j, {\rm min}} = 0.5 ^{\deg}$ for the FSRQ case or $\phi_{j, {\rm min}} = 0.001 ^{\deg}$ for the BLL case are consistent with a highly columnated jet if one assumes a conical shape. 

In the FSRQ case, the 2014 flare is nearly in equipartition $\zeta_{e+p} = 1$ between the energy density of the particles and the field (Table \ref{tbl-calculated}), where we consider both the electron and proton energies.  In the BLL case, the jet is particle dominated. It is relatively common for electron dominated emission models of blazars to suggest that the jet is slightly particle dominated, so the unusual case here is actually the FSRQ.

We cannot formally calculate the magnetization parameter in this type of model, but there is an upper limit enforced by the electron Larmor radius that will fit inside the emitting region \citep{lewis16,lewis19}.  The upper limit of the magnetization parameter indicates that acceleration via reconnection is unlikely or subdominant in both FSRQ and BLL scenarios since $\sigma_{\rm max} \not\gg 1$. (Note that reconnection is not included in the model.) All of the power produced in the jet through both dominant particle types and the magnetic field, is less than the luminosity of the jet $\mu_a < 1$, indicating that ordinary accretion \citep[e.g.][]{shakura73} is sufficient to power the jet in either case.

\begin{figure}[t]
\centering
\includegraphics[width=0.47\textwidth, trim={0 0 4 0}, clip]{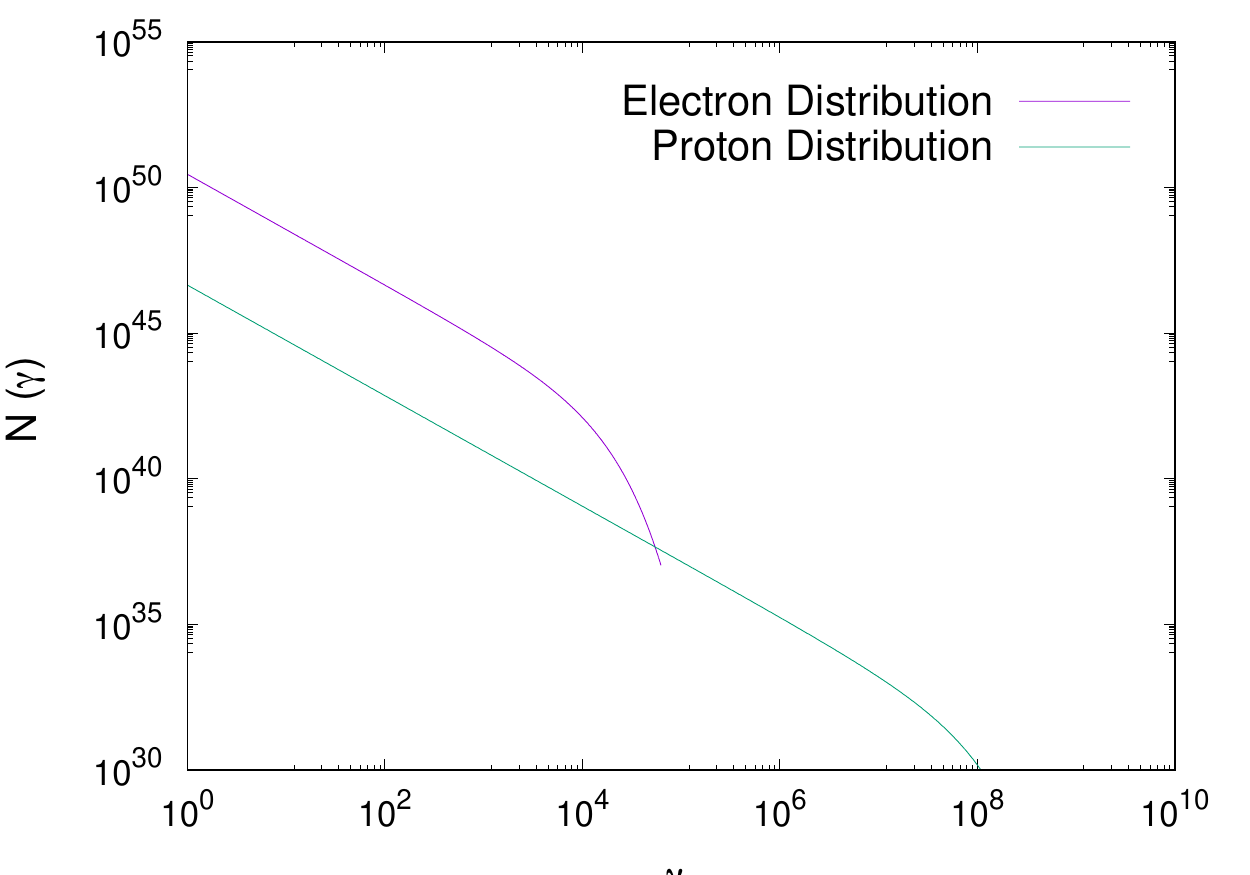}
\caption{Particle distributions for the BLL case with respect to the particle Lorentz factor $\g$. }
\label{fig-ED14ssc}
\end{figure}

We co-solve the particle transport equations independently for each set of parameters considered. Since the transport equation contains both acceleration and emission processes, these are reflected in the shape of the particle distribution. In particular the slope of the power-law is determined in this case predominantly by the ratio between the shock acceleration/adiabatic expansion and the stochastic acceleration. Since these parameters are consistent between the electron and proton equations, as implied by the assumption that they are co-spatial, the power-law slope is the same in both distributions (Figure \ref{fig-ED14ssc}).  The position of the exponential turnover in energy is determined by the rate of acceleration and the rate of cooling. Where the cooling rate becomes dominant, the particle spectrum will turn over \citep[e.g.][]{lewis19}. This occurs in different places for the electrons and the protons largely because of their mass difference, but it is also worth noting that the electrons lose energy to 3 processes (synchrotron, SSC, and EC) while the protons only cool through synchrotron in the current models. The rate of particles being injected into the jet is constrained to be equal for electrons and protons, so the total number of each particle type in the jet should be similar (although there is some conversion through $p\g$ cascades). There are significantly fewer protons at lower energies (where the electrons congregate) because there are a large number of protons at significantly higher energies (where there are no electrons).

\subsection{2017 Epoch}
\label{2017}

The broadband spectral data for the 2017 flare  \citep{Aartsen18} is distinct from the historical range in Section \ref{history} for this source in the $\g$-ray regime. So, it is expected that some parameters will vary more between the historical data and 2017 flare than between the historical data and 2014 flare.  The \citet{Aartsen18} reduction of the {\it Fermi}-LAT $\g$-ray data for this period uses a 28 day window centered on the neutrino detection, and their methodology gives a flattened spectrum. Alternatively, \citet{Keivani:2018rnh}'s analysis included approximately the same temporal cuts, but resulted in a more curved spectrum. It is not clear which data reduction method should be considered correct, but it is difficult to explain a flattened spectrum with a single dominant emission component since they tend to resemble log-parabolas. Similarly to Section \ref{2014}, we employ a FSRQ inspired transport equation model where the parameters are tuned to emphasize EC for the $\g$-rays.  Then, we examine the BLL interpretation with a transport model that treats the SSC cooling consistently for the electrons, and tune the free parameters to emphasize the SSC component in the $\g$-rays.

\begin{figure}[t]
\centering
\includegraphics[width=0.47\textwidth, trim={4 0 4 0}, clip]{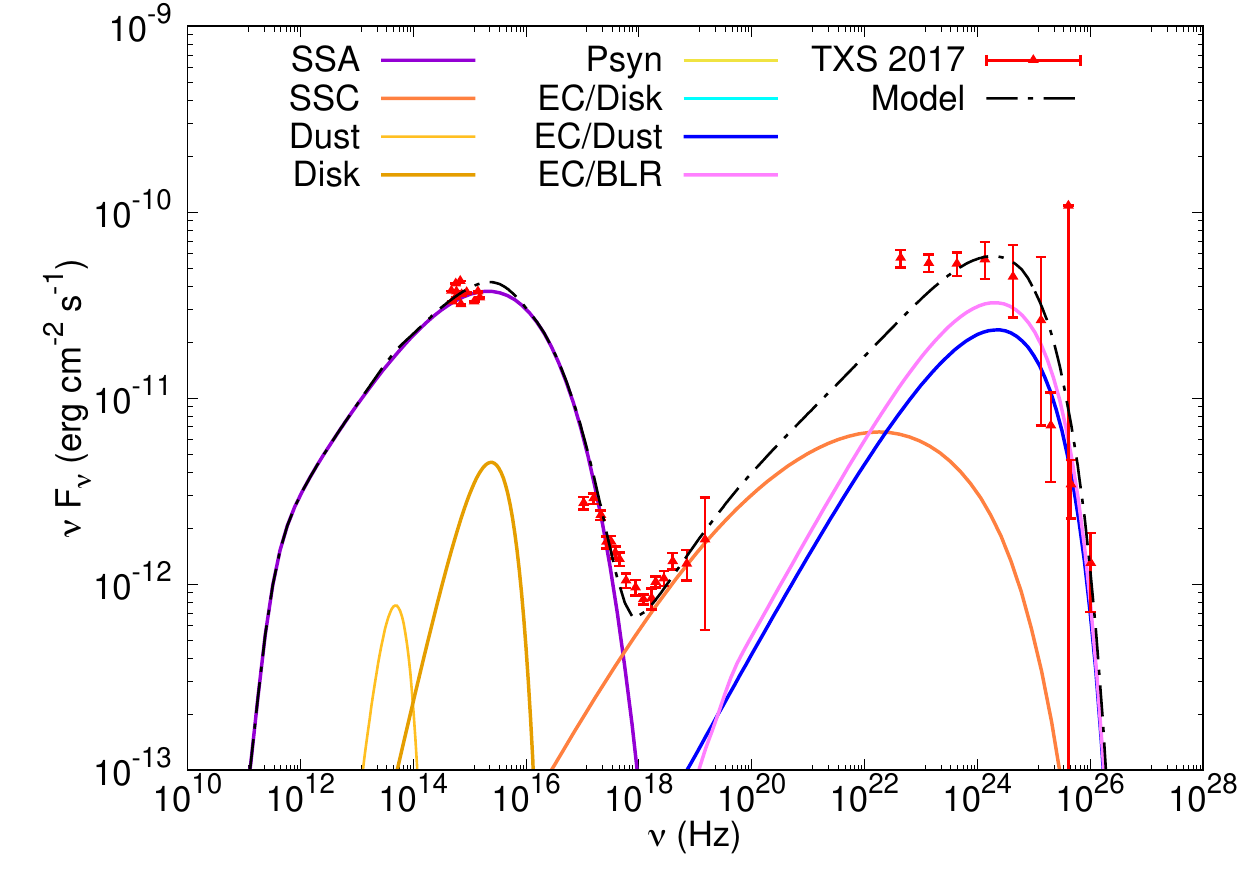}
\caption{Comparison of the FSRQ leptohadronic model to broadband multiwavelength data from the 2017 flare of TXS 0506+056 \citep[][in red]{Aartsen18}. The individual components of the emission model are labeled with the EC components dominating in the $\g$-rays. The co-added total emission from the model is given by the dot dashed line.}
\label{fig-flare17no}
\end{figure}

In Figure \ref{fig-flare17no}, we assume that TXS 0506+056 is a FSRQ. The optical data is well explained by the synchrotron peak. The X-ray regime is well covered and may be explained by the convergence of the synchrotron and SSC components around $\nu = 10^{18}$Hz.  The high-energy $\g$-ray data is well explained by the EC processes. However, it is not possible to recreate the lower-energy $\g$-ray spectrum simultaneously in this configuration. In particular, the EC components are too narrow to explain this version of the data.

\begin{figure}[t]
\centering
\includegraphics[width=0.47\textwidth, trim={4 0 4 0}, clip]{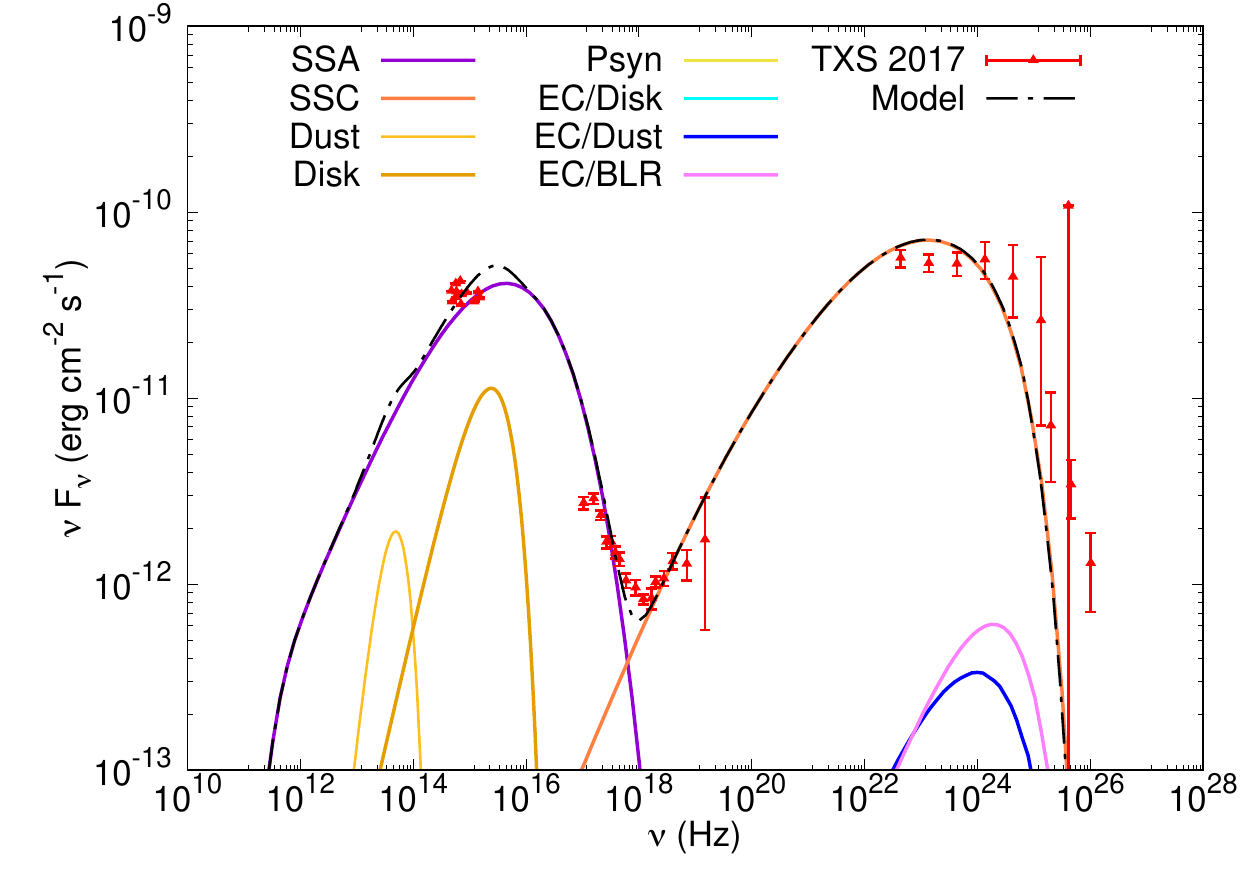}
\caption{Comparison of the BLL leptohadronic model to broadband multiwavelength data from the 2017 flare of TXS 0506+056 \citep[][in red]{Aartsen18}. The individual components of the emission model are labeled with the SSC components dominating in the $\g$-rays. The co-added total emission from the model is given by the dot dashed line.}
\label{fig-flare17ssc}
\end{figure}

Figure \ref{fig-flare17ssc} explores the same data set \citep{Aartsen18}, under the assumption that TXS 0506+056 is a BLL. In this case, the electron transport equation includes SSC as a cooling term, and the parameters are tuned intentionally to emphasize the SSC component rather than EC in the $\g$-ray regime. Here again, the optical data is well explained with electron synchrotron emission, with or without the illustrated flare of the accretion disk. The BLL model matches the X-ray data about as well as the FSRQ model in Figure \ref{fig-flare17no}. The fit to the lower end of the $\g$-ray spectrum is improved in this configuration, with the SSC component being broader. However, it is not sufficiently broad to simultaneously explain the X-rays and the higher frequency $\g$-ray points, even though the general shape is close. In order to suppress the EC/BLR, it was not necessary to move the emitting region much further from the outer edge of the BLR or to alter the dust reprocessing efficiency.  So, those parameters are similar to the FSRQ case (Tables \ref{tbl-freeparams} \& \ref{tbl-calculated}).  

The variability timescale ($t_{\rm var}$), signalling the size of the emission region ($R'_b$), for the FSRQ and BLL analyses of the 2017 flare are between the extremes that we explored in the analysis of the 2014 flare, with the BLL case being slightly larger (Table \ref{tbl-freeparams}).  The jet opening angle (in a conical approximation) is especially small for both models of the 2017 flare, which may indicate that the jet is very columnated or that the jet not well approximated by a cone out past the BLR, where the emission region seems to be located in this picture. Increasing the Doppler boosting (bulk Lorentz factor) is a way to broaden the predicted spectrum without increasing the particle acceleration. It is interesting that the Doppler factor is higher in the FSRQ case, since the EC component tends to reach higher frequencies than SSC.  However, the rate of particle acceleration is several orders of magnitude higher in the BLL case in order to fuel the the increased synchrotron cooling rate (higher magnetic field) and broader SSC component. 

Both the FSRQ and BLL cases explored indicate the jet is particle dominated during the 2017 flare, although moreso in the FSRQ case. Both cases also seem well explained by ordinary accretion since the jet power to accretion luminosity ratio ($\mu_a < 1$). The Compton dominance $A_C$ is larger for the 2017 flare than the 2014 flare or the historical data for this source. The value for Compton dominance in Table \ref{tbl-calculated} is is more accurate in the FSRQ interpretation since the calculation is based on the ratio between the synchrotron and EC immensities. Since EC is suppressed in the BLL case, the $\g$-ray luminosity is not represented by the luminosity of the EC component. The total jet luminosity is higher in the BLL case because the SSC component is broader - consistent with the 2014 flare analysis. It is especially interesting that the maximum magnetization parameter $\sigma_{\rm max} \gg 1$ in the BLL case. When the magnetization parameter is near or less than 1, stochastic acceleration is likely. Whereas, when the magnetization is much greater than 1, the system is in the reconnection regime. stochastic acceleration and reconnection generally do not happen simultaneously. It is worth mentioning that reconnection is not included in this family of models and that the actual magnetization parameter may not be near the maximum allowed. However, it would be interesting to followup the analysis with a model that includes reconnection to further explore the 2017 flare in the BLL interpretation.  

\begin{figure}[t]
\centering
\includegraphics[width=0.47\textwidth, trim={0 0 4 0}, clip]{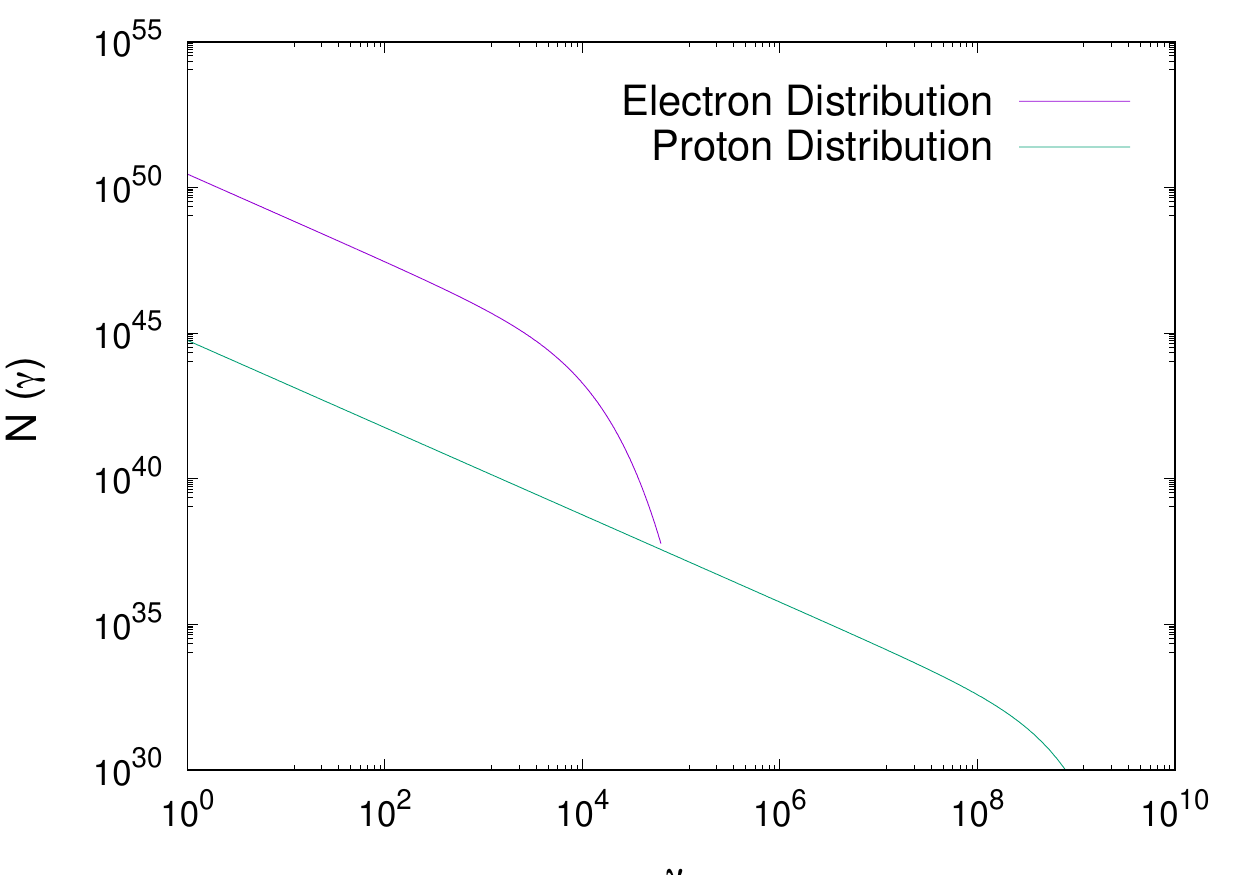}
\caption{Particle distributions for the BLL case with respect to the particle Lorentz factor $\g$. }
\label{fig-ED17ssc}
\end{figure}

Figure \ref{fig-ED17ssc} shows the electron and proton distributions with respect to the particle Lorentz factor for the BLL interpretation of the 2017 flare. In comparison to Figure \ref{fig-ED14ssc}, the electron distribution is qualitatively similar, but the slope of the power-law section is slightly less because the ratio between stochastic acceleration and shock acceleration/adiabatic expansion is closer to $a=-2$, which signifies horizontal \citep{lewis19}. The proton distribution power-law matches the electron distribution since the same acceleration rates apply to both populations. However, in comparing between the 2014 and 2017 particle distributions, the 2017 proton break energy is about an order of magnitude higher Lorentz factor since there is overall more acceleration available to the particles in the 2017 simulation. Since the proton population stretches to higher energies and more protons achieve higher energies, there are fewer protons that accumulate at lower energies and the electron and proton curves appear more separated.

\section{Simulations}
\label{Sims}

\color{black}
The Medium-Energy Gamma-ray Astronomy library (MEGAlib) software package\footnote{Available at \url{https://megalibtoolkit.com/home.html}} is a standard tool in MeV astronomy~\citep{2006NewAR..50..629Z}. MEGAlib was first developed for the MEGA instrument, and since then it has been further developed and successfully applied to a number of hard X-ray and $\gamma$-ray telescopes including, the Nuclear Compton Telescope (NCT), the Nuclear Spectroscopic Telescope Array (NuSTAR), the Imaging Compton Telescope (COMPTEL), and most recently, the Compton Spectrometer and Imager (COSI). Among other things, MEGAlib simulates the emission from a source, simulates the instrument response, performs the event reconstruction, and generates data for a given detector design, exposure time, and background emission (instrumental and astrophysical). Specifically, the source flux is generated from Monte Carlo simulations using \textit{cosima}, the event reconstruction is performed using \textit{revan}, and the high-level data analysis is done using \textit{mimrec}. MEGAlib is written in C++ and utilizes ROOT (v6.18.04) and Geant4 (v10.02.p03).


We use MEGAlib (v3.02.00) to simulate AMEGO-X observations of the source TXS 0506+056 during the two flare periods. The simulations employ the latest version of the AMEGO-X detector, as described in \citet{2021arXiv210802860F}. Events are simulated between 100 keV to 1 GeV, using 10 energy bins. For the first neutrino event we use a total flare time of 158 days \citep{IceCube:2018cha}, and for the second neutrino event we use a total flare time of 120 days \citep{IceCube:2018dnn}. To account for the AMEGO-X duty cycle, the exposure time is taken to be 20\% of the flare time. For simplicity, we simulate the source on axis, and perform just a single simulation. This will be updated in the next version of the draft to include 1000 simulations and the average AMEGO-X off-axis angle for a source of $\sim$30$^\circ$.  

\begin{figure}[t]
\centering
\includegraphics[width=0.48\textwidth]{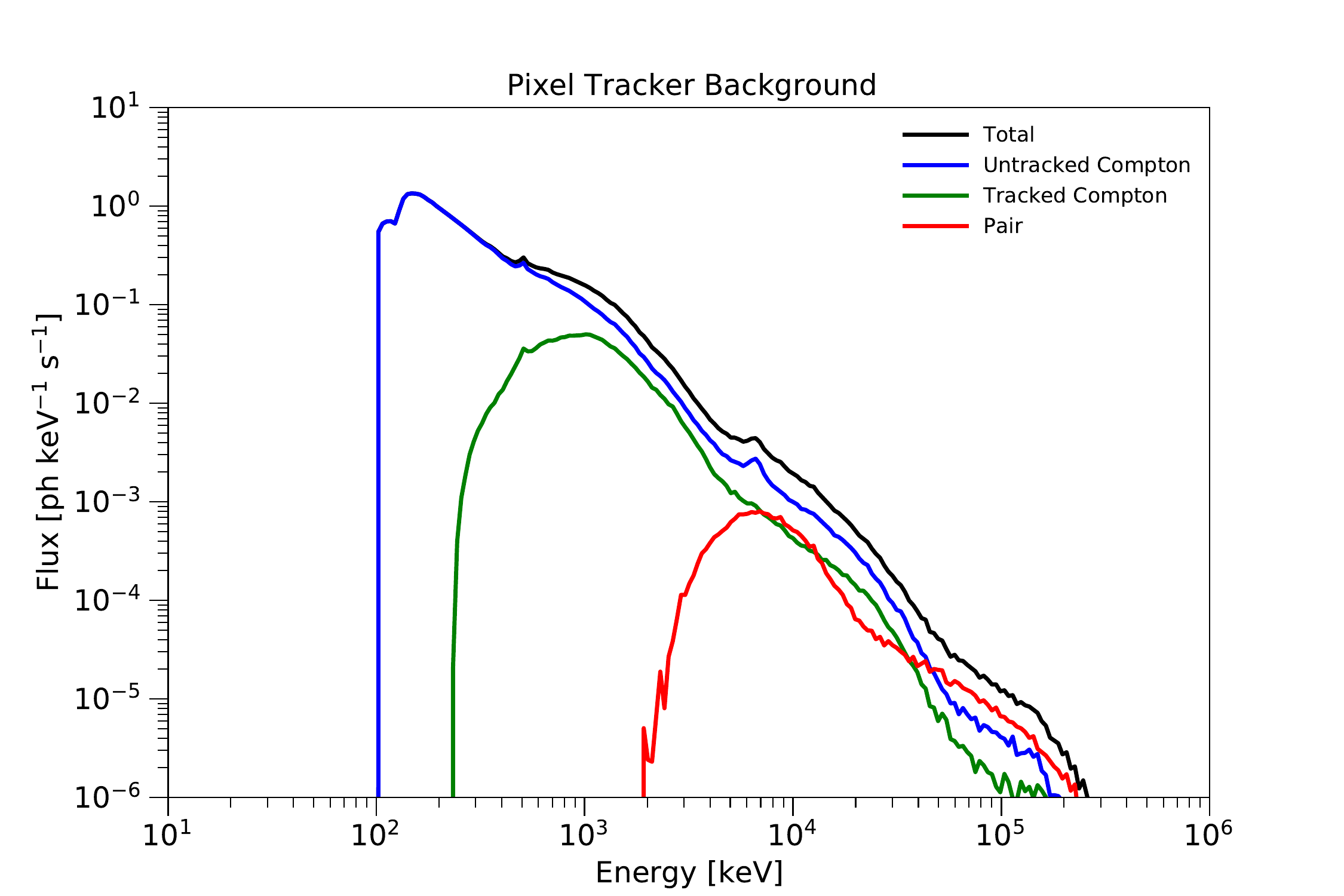}
\caption{AMEGO-X background rates (including both astrophysical and instrumental contributions). The background is comprised of (from left to right) untracked Compton events (blue), tracked Compton events (green), and pair events (red). The total background is the sum of the three components and is shown with the black curve.} 
\label{fig:BG}
\end{figure}

Simulating the background emission is very computationally expensive, and currently it is not practical to obtain simulated backgrounds for the full exposure time. We have therefore developed an analysis pipeline\footnote{Available at~\url{https://github.com/ckarwin/AMEGO_X_Simulations}} that enables us to simulate the source and background separately, and then combine the respective counts in order to obtain the signal-to-noise (SN) and statistical error ($\sigma$) on the flux, defined as
\begin{equation}
    SN = \frac{S}{\sqrt{S+B}},
\end{equation}

\begin{equation}
    \sigma = \sqrt{S+B},
\end{equation}
where $S$ is the source counts and $B$ is the background counts. The background is simulated for 1 hr and scaled up to the observation time. It includes both instrumental and astrophysical components. There are contributions from hadrons, hadronic decay, leptons, photons, and trapped hadronic decay. Note that the prompt trapped hadronic and trapped leptonic components contribute mainly during passage of the Southern Atlantic Anomaly (SAA) when the detector is expected to be off, and thus we do not include them in the overall background. It should also be noted that the current AMEGO-X background model includes the extragalactic diffuse emission from \citet{1999ApJ...520..124G}, but it does not include the Galactic diffuse component. Fig.~\ref{fig:BG} shows the total background rate, as well as the rates for the three constituent event types: untracked Compton, tracked Compton, and pair.

The counts spectrum (C) is output from \textit{mimrec} (in units of counts/keV). This is converted to photon flux ($dN/dE$) using the effective area ($A_{\mathrm{eff}}$) and exposure time (t):

\begin{equation}
\frac{dN}{dE} = \frac{C}{A_{\mathrm{eff}} \times t},
\end{equation}

\begin{equation}
A_{\mathrm{eff}} = \frac{N_{\mathrm{obs}}}{N_{\mathrm{sim}}} \times A_{\mathrm{sph}},
\end{equation}
where $N_\mathrm{obs}$ and $N_\mathrm{sim}$ are the number of observed and simulated counts, respectively, and $A_{\mathrm{sph}}$ is the area of the surrounding sphere from which the simulated events are launched (see MEGAlib documentation for more details).  

To optimize the detection sensitivity the pipeline utilizes an energy-dependent extraction region for both the source and background counts. The region is determined by the energy-dependent angular resolution, as given in~\citet{2021arXiv210802860F}, and is set self-consistently for source and background. Note that in the Compton regime the angular resolution is a measure of the full-width-at-half-maximum of the ARM (angular resolution measure), and in the pair regime the angular resolution is a measure of the 95\% containment angle. Additionally, we also optimize with respect to event type, i.e.~for each energy bin we consider the SN separately for untracked Compton, tracked Compton, and pair events, selecting the case that gives the highest SN. 


Since AMEGO-X is a highly versatile instrument, covering both the Compton and pair regime, we report measurements for two distinct energy regimes, which we refer to as the low energy band (extending from 100 keV to 2.3 MeV) and the high energy band (extending from 15.2 MeV to 432.9 MeV). Note that the specific ranges of the bands were choosen to coincide with the SED binning. Below $\sim$3 MeV most of the events are from Compton interactions, whereas above $\sim$3 MeV most of the events are from pair interactions. We define the high energy band to start a bit above 3 MeV in order to avoid the transition region which has a lower sensitivity. Note that above $\sim$400 MeV the effective area quickly falls off and there are no counts in our simulations. 

\begin{figure}[t]
\centering
\includegraphics[width=0.48\textwidth]{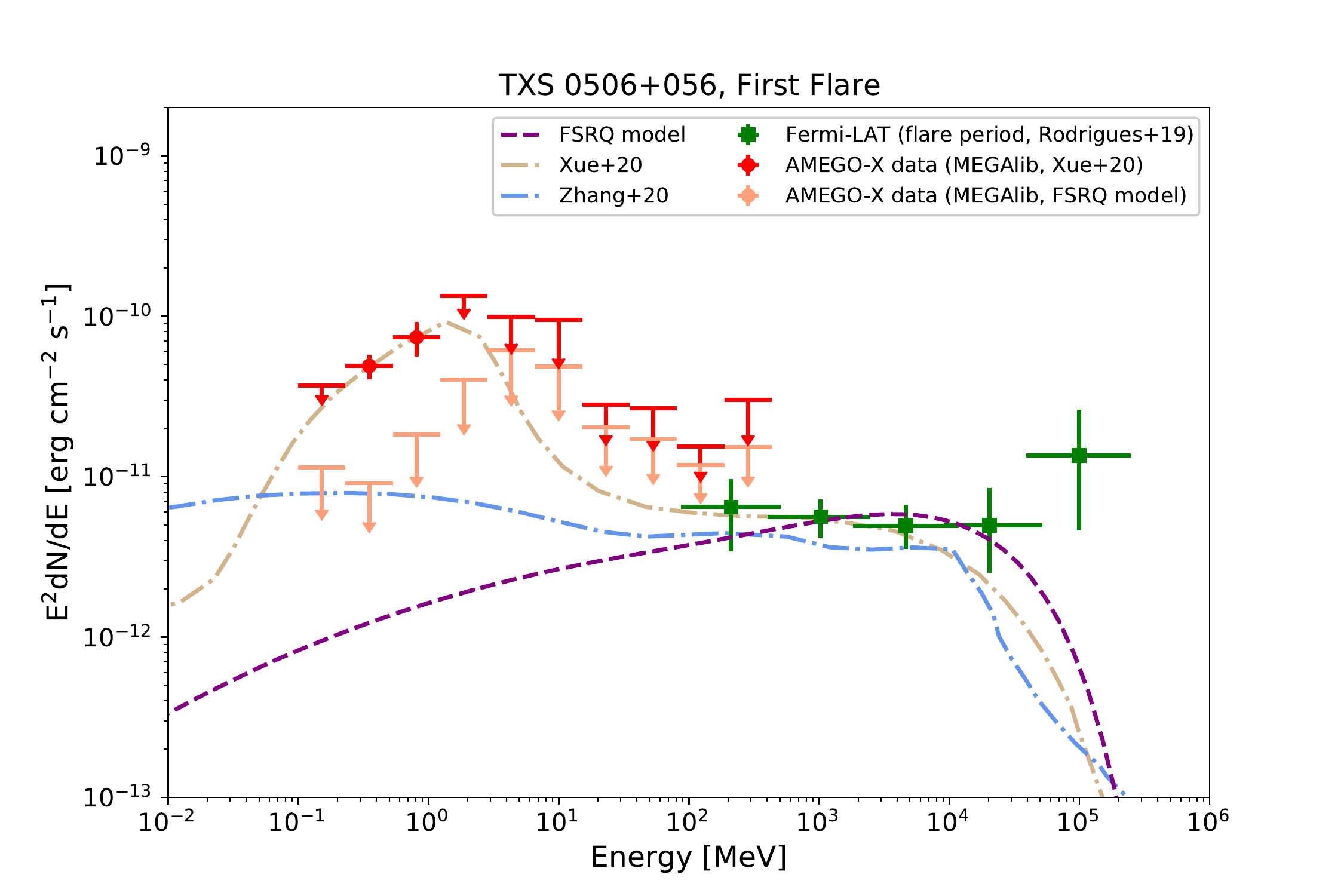}
\caption{Expected AMEGO-X flux for the first flare period. The FSRQ model is shown with the purple dashed curve, and the simulated data is shown with peach markers. The model from~\citet{2021ApJ...906...51X} is shown with the tan dash-dot curve, and the simulated data is shown with red markers. For reference, the blue dash-dot curve shows the predicted flux from~\citet{2020ApJ...889..118Z}. The green markers show the LAT date during the flare period from~\citet{2019ApJ...874L..29R}.} 
\label{fig:SED1}
\end{figure}

The simulated SED for the first event is shown in Fig.~\ref{fig:SED1}. Our representative FSRQ model is shown with the dashed purple curve, and the corresponding simulated data is shown with the peach markers. As can be seen, the emission falls below the AMEGO-X sensitivity for the entire energy range. The upper limits are plotted as the flux + 1$\sigma$ error, and thus they represent the 68\% confidence level. We also simulate the predicted flux from~\citet{2021ApJ...906...51X}. The model is shown with the tan dash-dot curve, and the corresponding data is shown with the red markers. In this case AMEGO-X is sensitive in the low energy band. In particular, the integrated flux in the low energy band is $(23.8 \pm 5.0) \times10^{-11} \mathrm{\ erg \ cm^{-2} \ s^{-1}}$, with a corresponding SN=7.8. The upper limit in the high energy band is $6.5\times10^{-11} \mathrm{\ erg \ cm^{-2} \ s^{-1}}$. For comparison, Fig.~\ref{fig:SED1} also shows the \textit{Fermi}-LAT data for the flare period. 

For the second flare period we simulate just the FSRQ model. Note that this event did not have efficient neutrino production, and thus an enhanced flux in the MeV band is not expected (e.g.~\citet{2021ApJ...906...51X}). In this case the high energy band is significantly detected, with a measured flux of $(8.9 \pm 2.9) \times10^{-11} \mathrm{\ erg \ cm^{-2} \ s^{-1}}$ and a corresponding SN = 3.5. In the low energy band we measure an upper limit of $7.5\times10^{-11} \mathrm{\ erg \ cm^{-2} \ s^{-1}}$. 

\begin{figure}
\centering
\includegraphics[width=0.48\textwidth]{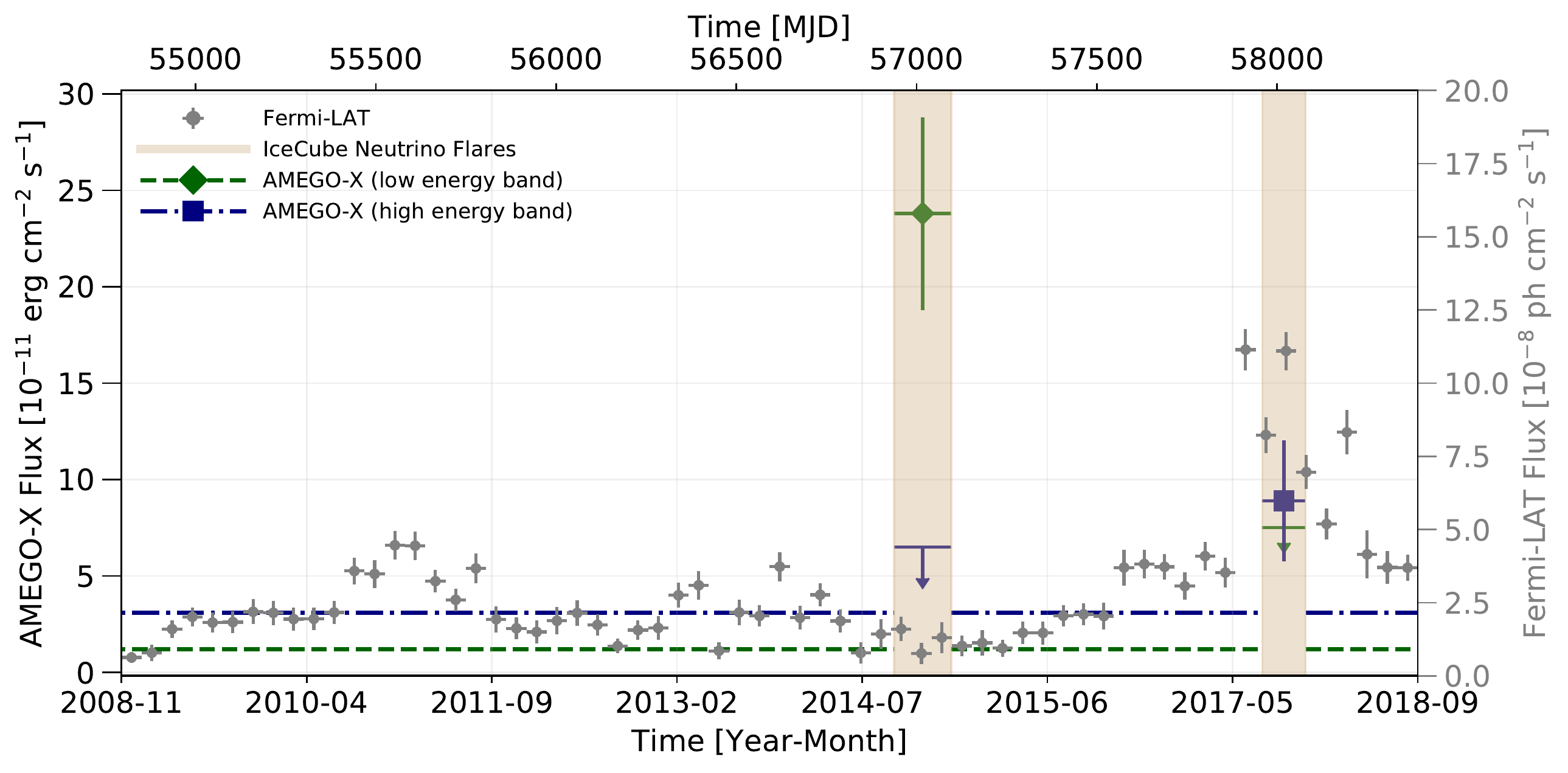}
\caption{TXS 0506+056 $\gamma$-ray light curve covering roughly 10 years of the LAT mission. The tan bands show the periods of the two neutrino flares. The y-axis on the left gives the AMEGO-X flux, which is plotted for both the low and high energy bands. The green dash and purple dash-dot lines give an estimate of the quiescent flux. The LAT data is shown with grey (from~\citet{Petropoulou:2019zqp}) and corresponds to the y-axis on the right.} 
\label{fig:LC}
\end{figure}

Fig.~\ref{fig:LC} shows the $\gamma$-ray light curve of TXS 0506+056 over roughly a 10 year observational period. The tan bands correspond to the two flare periods, and the simulated AMEGO-X data is plotted for both the high and low energy bands, corresponding to the y-axis on the left. For the first event we show the simulated data corresponding to the model from~\citet{2021ApJ...906...51X}. The LAT data is shown with grey markers, corresponding to the y-axis on the right. To estimate the quiescent state flux we scale the FSRQ model to the LAT data for the full mission. The integrated flux is $1.2\times10^{-11} \mathrm{\ erg \ cm^{-2} \ s^{-1}}$ and $3.1\times10^{-11} \mathrm{\ erg \ cm^{-2} \ s^{-1}}$ in the low energy band and high energy band, respectively. As can be seen, for a cascade scenario as described in~\citet{2021ApJ...906...51X}, AMEGO-X would have been able to significantly detect the first event (in the low energy band), which was not detected by the LAT.

\section{Discussion \& Conclusions}
\label{Disc}

In Section \ref{Data}, we compared the effect of different assumptions about TXS 0506+056 on the fits to specific data sets and the physical parameter regimes that produce those simulated spectra. In particular, we categorically assumed that the full multiwavelength spectrum is produced in a single zone that is homogeneous, the peak of a flare may be approximated by a steady state, the acceleration and emission regions are co-spatial, and that the injection rate of protons and electrons is the same. In the case of the FSRQ interpretation, we additionally constrain that the EC emission dominates the SSC at high energies, and in turn for the BLL cases that the SSC emission dominates EC. Most of these assumptions stand in contrast to existing literature for this source in one way or another, and it is informative to explore different possible scenarios systematically. Since many of the assumptions about the physical picture disagree, it can be difficult to make direct comparisons between parameters.

The accretion disk luminosity used in this work is somewhat larger than that used in \citet{Xue21} based on extrapolations from references therein. The accretion disk emission curve is under the peak synchrotron emission curve in both sets of models, so it is not directly important to the fits, but the value can change our perception of the energy balance between the accretion disk and the jet. 

\citet{Xue21} demonstrate that the presence of a corona for a 2 zone model, is able to generate the neutrino signature, while avoiding overproduction of photon emission because the same corona photon field that generates sufficient energy density for the neutrino production also creates the opacity that blocks spectral emission.  Since the presence of an X-ray corona is not observationaly confirmed, it is neglected in this work. The emitting regions we consider across scenarios are between where \citet{Xue21} predict their inner and outer blobs would occur. Similarly the size of our blobs are between the sizes predicted of inner and outer blobs in \citet{Xue21}. The magnetic field ranges between the two works are similar.  This work tends to find higher Doppler factors. However, from a modeling perspective Doppler brightening can produce effects on the spectrum that are in some ways similar to the effects of acceleration mechanisms. The acceleration of particles here is treated as co-spatial with the primary cooling and includes both first and second-order Fermi processes. As is common practice, \citet{Xue21} seem to use a particle injection spectrum that is pre-accelerated, and usually interpreted as shock acceleration. 

In examining the 2017 flare, \citet{Keivani18} generate fits to the data using a single zone model with EC as the dominant high-energy bump, where the hadronic component that produces neutrinos is subdominant. This physical picture is more comparable to the work presented here. While their electron spectrum is qualitatively similar to that derived here, they do not treat acceleration as co-spatial with the emission region. The maximum proton energies are also similar despite the different methods, reinforcing the idea that the shape of the particle spectrum is important to the spectra produced, meaning that spectral models that match the same data set will require similar underlying particle spectra. This further implies that something  about particle acceleration may be discerned from complete, simultaneous multiwavelength spectra, and the addition of neutrino spectra provides further constraints. 

With regard to their EC leptonic model, \citet{Keivani18} use a BLR size similar to our finding of the position of the Lyman-$\alpha$ line in the BLR, which is often used as a proxy for the entire BLR because it has the highest intrinsic energy density of any line in the BLR. However, we know from reverberation mapping \citep[e.g.][]{kaspi05,kaspi07,finke16} that BLR regions are predictably stratified due to a photoionization gradient outside of the accretion disk. If one considers the relativistic motion of the emission region the Lyman-$\alpha$ line may not have the highest energy density in the comoving frame from the blob location. Therefore, it is beneficial to treat multiple lines when calculation EC/BLR and to use one of the expected outer lines (e.g. H$\beta$) when approximating the extent of the BLR. 

\citet{Keivani18} also consider a leptohadronic modes in which cascade pair production and proton synchrotron produce the high-energy bump, a configuration not currently examined here. In those simulation, they find significantly higher magnetic field values, which is to be expected in that picture. The neutrino production in the proton dominated leptohadronic model of \citet{Keivani18} is sensitive to the Doppler boosting, and is one of the better representations of the full spectrum for the 2017 flare in the literature, suggesting that a good extension of this work would be to emphasize hadronic emission processes. 

Where the models explained in this work are comparable with the extensive literature on this source, there is broad agreement. However, the breadth of approaches employed makes direct comparisons difficult.

\subsection{Simulations \& Implications}
\label{imply}

Some of the challenges in constraining models of TXS 0506+056 arise from the incomplete multiwavelength spectra available. Most notably, there is currently a large sensitivity gap in the MeV band, yet this band may be critical for detecting the electromagnetic counterpart of neutrino flares. In particular, the counterparts to bright neutrino emission will not be detected by high-energy gamma-ray telescopes such as the LAT, since the radiation fields required for efficient neutrino production make the source opaque to high-energy gamma rays. Even in the case of the 2017 co-detection of neutrinos with a GeV flare, it is still not possible to definitively state whether something like a corona could contribute to the neutrino production. As shown in Section~\ref{Sims}, during the 2014-2015 TXS 0506+056 orphan neutrino flare, an MeV instrument such as AMEGO-X would have been able to detect a significant MeV flare even though there was no detectable GeV flare, for a model similar to that described in~\citet{Xue21}. Observational evidence of coronae in jetted AGN would be a significant discovery, and illuminate answers to long debated questions, including how jets form and the source of astrophysical neutrinos.

\begin{acknowledgments}
T.L. is supported by an appointment to the NASA Postdoctoral Program at NASA Goddard Space Flight Center, administered by Universities Space Research Association under contract with NASA.  H.F. acknowledges support by NASA under award number 80GSFC21M0002. The authors are pleased to acknowledge conversations with Marco Ajello, as well as the AMEGO-X team. 
\end{acknowledgments}

\bibliographystyle{aasjournal}
\bibliography{ms}

\end{document}